\magnification=\magstep1
\tolerance=500
\bigskip
\rightline{18 November, 2020}
\centerline{\bf Spin and Entanglement in General Relativity}
\bigskip
\centerline{Lawrence P. Horwitz}
\bigskip
 \centerline{ Tel Aviv University, Ramat Aviv, 69978 Israel}
\centerline{ Ariel University, Ariel, 40700 Israel}
\centerline{ Bar Ilan University, Ramat Gan, 52900 Israel}
\bigskip
\noindent{\it Abstract}
\bigskip
\par In a previous paper, we have shown how the classical and quantum relativistic dynamics of the Stueckelberg-Horwitz-Piron [SHP] theory can be embedded in general relativity (GR). We briefly review the SHP theory here and, in particular, the formulation of the theory of spin in the framework of relativistic quantum theory. We show here how the quantum theory of relativistic spin can be embedded, using a theorem of Abraham, Marsden and Ratiu, and also explicit derivation, into the framework of GR by constructing a local induced representation. The relation to the work of Fock and Ivanenko is also discussed. We show that in a gravitational field there is a highly complex structure for the spin distribution in the support of the wave function. We then discuss entanglement for the spins in a two body system.
\bigskip
\bigskip
\noindent{\it 1. Introduction}
\bigskip
\par The relativistic canonical Hamiltonian dynamics of Stueckelberg, Horwitz and Piron (SHP)[1] with scalar potential and gauge field interactions for single and many body theory can, by local coordinate transformation, be embedded into the framework of general relativity (GR)[2][3] (to be called SHPGR). We first review the structure of this embedding and then discuss the introduction of spin, angular momentum and entanglement in this framework.
\par The theory was originally formulated for a single event (associated by its world line with a particle) by Stueckelberg in 1941[4][5][6]. Stueckelberg  envisaged the motion of an event along a world line in spacetime that can curve, due to interaction, and turn to flow backward in time, resulting in the phenomenon of pair annihilation in classical dynamics.  Since this world line is not single valued in $t$, Stueckelberg parametrized it by an invariant monotonic parameter $\tau$. 
\par The theory was generalized by Horwitz and Piron in 1973 [7] to be applicable to many body systems by assuming that the parameter $\tau$ is {\it universal} (as for Newtonian time [8][9]), enabling them to solve the two body central potential problem classically. A solution for the quantum case was found later by Arshanksy and Horwitz [10][11][12], both for bound states and scattering theory, for interaction represented by a central potential function. 
 \par Performing local coordinate transformations from the flat tangent space, for which we label coordinates and momenta $\xi^\mu$ and $\pi_\mu$, to coordinates on a general manifold, which we label $x^\mu$, along with the corresponding transformation of the momenta (on the cotangent space of the original Minkowski manifold), which we label $p_\mu$, one obtains the SHP theory in the curved space of general coordinates and momenta with a canonical Hamilton-Lagrange (symplectic)[2][3] structure. We shall refer to this generalization as SHPGR.
 \par The invariance of the Poisson bracket\footnote{*}{We provide an explicit proof of the invariance of the Poisson bracket in connection with our discussion of the spin in a later section.} under local coordinate transformations provides a basis for the canonical quantization of the theory, for which the evolution under $\tau$ is determined by the Lorentz covariant form of the Stueckelberg-Schr\"odinger equation [1][4].   
\par This method was applied also to the many body case [13], for which the SHP Hamiltonian is a sum of terms quadratic in four momentum with a many body potential term. Each particle moves {\it locally} tangentially infinitesimally close to a flat Minkowski space, the tangent space of the general manifold of motions at that point; these local motions can then be mapped at each point $x^\mu$ by coordinate transformation into the curvilinear coordinates in that neighborhood reflecting the curvature induced by the Einstein equations.
\par We assume a $\tau$ independent
background gravitational field; the local coordinate transformations from the flat Minkowski space to the curved space are taken to be independent of $\tau$, consistently with an energy momentum tensor that is $\tau$ independent. In a more dynamical setting, when the energy momentum tensor depends on $\tau$, the spacetime evolves nontrivially;
the transformations from the local Minkowski coordinates to the curved space coordinates then depend on $\tau$; this situtation was discussed in [2]and by Land[14], but will not be treated here.
\par The theory of intrinsic angular momentum of a particle in the framework of relativistic quantum theory for special relativity was worked out by Horwitz and Arshansky [12]
(see also [1]) following the method of induced representations of Wigner [15], for which the representation was induced as the stability subgroup leaving the four-momentum invariant, but instead using a timelike vector $n^\mu$, transforming with the Lorentz group and independent of momentum.  The necessity for this is that the action of the Lorentz group on the induced representation of the angular momentum depends on the inducing vector. When computing the expectation value of $\xi^\mu$, represented (in momentum space) by $i{\partial \over \partial \pi_\mu}$ in the relativistic quantum theory, as discussed below, this derivative would destroy the unitarity of a representation induced on $\pi^\mu$. This expectation value would then not transform as a vector under the Lorentz group.
 \par The generators of the Lorentz group acting both on $\{\xi^\mu\}$ and $\{n^\mu\}$, in the relativistic quantum theory [1][12], are
 $$\eqalign{M^{\mu \nu} &= \xi^\mu \pi^\nu  -\xi^\nu \pi^\nu\cr
 &-i \bigl(n^\mu {\partial \over \partial n_\nu}-n^\nu {\partial \over \partial n_\mu}\bigr),\cr}
 \eqno(1.1)$$
 where indices are raised and lowered by the Minkowski metric $\eta_{\mu\nu}= (-1,+1,+1,+1)$. Under the action of the group generated by this set of operators,  $M^{\mu \nu}$ is a Lorentz tensor.  By the equivalence principle, the Lorentz group acts in the locally flat freely falling frame (tangent space). It is therefore essential for the embedding of the special relativistic theory into GR that the set of local generators transform under the local embedding diffeomorphisms
as covariant tensors.  We shall see that this follows by an isomorphism theorem of Abraham, Marsden and Ratui [16] and by explicit calculation.
\par In the following we briefly review the SHP theory and its imbedding into the curved space of GR [2][3](SHPGR) for a single spinless particle (the many  body case in Minkowski space was treated in ref. [13]) and then turn to discuss the representations of a particle with spin in GR. We shall see that, in the presence of a gravitatioal field, due to the properties of parallel transport, the spin structure of the wave function is highly complex.  We then treat the manifestation of long range correlations in GR resulting in spin entanglement . 
\par
\bigskip
\noindent{\it 2. Classical Theory for a Single Particle in an External Potential}
\bigskip
\par In order to carry out the embedding of the SHP theory into the manifold of general relativity, we start by writing the SHP theory in a local tangent space, and carry out local coordinate transformations, following Einstein's use of the equivalence principle. 
\par We write the SHP Hamiltonian [1] as
$$K= {1\over 2M} \eta^{\mu\nu} \pi_\mu \pi_\nu + V(\xi) \eqno(2.1)$$
where $\eta^{\mu\nu}$ is the flat Minkowski metric $(-+++)$ and $\pi_\mu, \xi^\mu$ are the spacetime canonical momenta and coordinates in the local tangent space.
\par The existence of a potential term (which may be a Lorentz scalar), representing non-gravitational forces, implies that the ``free fall'' condition is replaced by a local dynamics carried along by the free falling system (an additional force acting on the particle within  the ``elevator'' according to the coordinates in the tangent space). 
\par The canonical equations are
$$ {\dot \xi}^\mu = {\partial K \over \partial \pi_\mu} \ \ \ \ \ \ {\dot \pi}_\mu =- {\partial K\over \partial \xi^\mu} = -{\partial V\over \partial \xi^\mu}, \eqno(2.2)$$
where the dot here indicates ${d\over d\tau}$, with $\tau$ the invariant universal ``world time''.
Since
$$\eqalign{{\dot \xi}^\mu &= {1 \over M} \eta^{\mu \nu} \pi_\nu,\cr
 {\rm or} \ \ \ \pi_\nu &=\eta_{\nu \mu} M{\dot \xi}^\mu, \cr}
       \eqno(2.3)$$
the Hamiltonian can then be written as\footnote{*}{ Note that, as clear from $(2.3)$, that ${\dot \xi}^0= {dt\over d\tau}$ has a sign {\it opposite} to $\pi_0$ which lies in the cotangent space of the manifold. The energy of the particle for a normal time-like particle should be positive (negative energy would correspond to an antiparticle [4][17]). The {\it physical momenta and energy} therefore correspond to the mapping $ \pi^\mu = \eta^{\mu\nu}\pi_\mu.$
back to the tangent space.}
$$K= {M\over 2} \eta_{\mu\nu}{\dot \xi}^\mu {\dot\xi}^\nu + V(\xi). \eqno(2.4)$$
 \par We now transform the local coordinates (contravariantly) according to the diffeomorphism
$$ d\xi^\mu = {\partial \xi^\mu \over \partial x^\lambda} dx^\lambda \eqno(2.5)$$
to relate small changes in $\xi$ to corresponding small changes in the coordinates $x$ on the curved space, so that
$$ {\dot \xi}^\mu = {\partial \xi^\mu \over \partial x^\lambda} {\dot x}^\lambda . \eqno(2.6)$$
The Hamiltonian then becomes
$$K =  {M\over 2} g_{\mu\nu}{\dot x}^\mu {\dot x}^\nu + V(x), \eqno(2.7)$$
where $V(x)$ is the potential at the point $x$ corresponding to the point $\xi$ in the tangent space that we have been considering,  and
$$ g_{\mu\nu}= \eta_{\lambda \sigma}
 {\partial \xi^\lambda \over \partial x^\mu}{\partial \xi^\sigma \over \partial x^\nu} \eqno(2.8)$$ 
Since $V(x)$ has the dimension of mass, one can think of this function as a scalar mass field, inducing forces acting in the local tangent space at each point. It may play the role of ``dark energy'' [18][19]. 
\par The corresponding Lagrangian in the curved space is then
$$L =  {M\over 2} g_{\mu\nu}{\dot x}^\mu {\dot x}^\nu - V(x), \eqno(2.9)$$
\par In the locally flat coordinates in the neighborhood of $x^\mu$, the symplectic structure of Hamiltonian mechanics ( {\it e.g.} da Silva [20] ) implies that the momentum    $\pi_\mu$\footnote{*}{We shall call the quantity $\pi_\mu$ in the cotangent space a {\it canonical momentum}, although it must be understood that its map back to the tangent space $\pi^\mu$ corresponds to the actual physically measureable energy  momentum.}, lying in the cotangent space of the manifold $\{\xi^\mu \}$, transforms covariantly under the local transformation $(1.5)$, {\it i.e.}, as does ${\partial \over \partial \xi^\mu}$, so that we may define
$$ p_\mu = {\partial \xi^\lambda \over \partial x^\mu} \pi_\lambda. \eqno(2.10)$$
This definition is consistent with the transformation properties of the momentum defined by the Lagrangian $(2.9)$:
$$ p_\mu = {\partial L(x,{\dot x}) \over \partial {\dot x}^\mu}, \eqno(2.11)$$
yielding
$$ p_\mu = M  g_{\mu\nu}{\dot x}^\nu. \eqno(2.12)$$
\par The second factor in the definition $(2.8)$ of $g_{\mu\nu}$ in $(2.12)$ acts on ${\dot x}^\nu$; with $(2.6)$ we then have (as in $(2.10)$)
$$\eqalign{ p_\mu &= M  \eta_{\lambda \sigma} {\partial \xi^\lambda \over \partial x^\mu}{\dot \xi}^\sigma\cr
&= {\partial \xi^\lambda \over \partial x^\mu} \pi_\lambda . \cr} \eqno(2.13)$$
\par As we have remarked above for the locally flat space, the {\it physical} energy and momenta are given, according
to the mapping,
$$ p^\mu = g^{\mu\nu} p_\nu = M {\dot x}^\nu \eqno(2.14) $$
back to the tangent space of the manifold, which also follows directly from the local coordinate transformation of $(2.3)$.
\par It is therefore evident from $(2.14)$ that
$$ {\dot p}^\mu = M {\ddot x}^\mu.  \eqno(2.15)$$
\par We see that ${\dot p}^\mu$, which should be interpreted as the force acting on the particle, is proportional to the {\it acceleration along the orbit of motion}. From the coordinate transformation of
$${\ddot \xi}^\mu = - {1 \over M} {\dot \pi}_\mu = - {1 \over M}\eta^{\mu \nu} {\partial V(\xi) \over \partial \xi^\nu} , \eqno(2.16)$$
one finds that $(2.15)$ is equivalent to the usual geodesic formula with an additional contribution due to the potential [2]
$$ {\dot p}^\mu =  M {\ddot x}^\nu=  -M {\Gamma^\sigma}_{\lambda \gamma} {\dot x}^\gamma {\dot x}^\lambda
  - g^{\sigma \lambda} {\partial V(x)\over\partial x^\lambda} \eqno(2.17)$$
\par The procedure that we have carried out here provides a canonical dynamical structure for the motions in the curvilinear coordinates. The Poisson bracket remains valid for the coordinates $\{x,p\}$[2], as we show explicitly below, so that
$$ [x^\mu, p_\nu]_{PB} = {\delta ^\mu}_\nu. \eqno(2.18)$$
\par One also finds that 
$$ [p_\mu, F(x)]_{PB} = - {\partial F \over \partial x^\mu}, \eqno(2.19)$$
so that $p_\mu$ acts infinitesimally as the generator of translation {\it along the coordinate curves} (in a geodesically complete manifold these may be taken to be geodesic curves) and  
$$[x^\mu, F(p)]_{PB} =  {\partial F(p) \over \partial p_\mu}, \eqno(2.20)$$
so that $x^\mu$ is the generator of translations in $p_\mu$.
\par The Poisson bracket structure of the mapping of the SHP theory into GR gives us a basis for quantization, following Dirac [17]. Properties of this quantum formulation were discussed in [2]. We now turn to the treatment of a particle with spin in this framework.
\vfill\eject
\bigskip
\noindent{\it 3. Quantum Theory}
\bigskip
\par Our discussion so far has been primarily classical. The Poisson brackets, as mentioned above, provide a  basis [2] for the corresponding qauntum theory on the curved space as discussed in SHPGR, for which the canonical commutation relations are
$$[x^\mu, p_\nu] = i\hbar{\delta^\mu}_\nu, \eqno(3.1)$$
as well as
$$[x^\mu, p^\nu] = i\hbar g^{\mu\nu}(x), \eqno(3.2)$$
and therefore
$$ [p_\mu, F(x)] = -i\hbar {\partial F \over \partial x^\mu} \eqno(3.3)$$
and
$$[x^\mu, F(p)] = i\hbar {\partial F(p) \over \partial p_\mu}. \eqno(3.4)$$
\par The scalar product for wave functions in the Hilbert space $L^2 (R^4,\sqrt{g} d^4 x)$ , with $g(x) = -det g^{\mu\nu}$ is defined (on this local diffeomorphism invariant measure) as [2][3]
$$(\psi, \chi) = \int d^4 x \sqrt{g}{\psi^*}_\tau (x) \chi_\tau (x).
\eqno(3.5)$$
\par The operator $-i\hbar {\partial/\partial x^\mu}$ is not Hermitian in this scalar product. However, 
$$ p_\mu = -i {\partial \over \partial x^\mu} - {i \over 2}{1 \over \sqrt{g(x)}}{\partial \over \partial x^\mu} \sqrt{g(x)}  \eqno(3.6)$$
is self-adjoint (somewhat in analogy to the Newton-Wigner position operator[21] in momentum space in Klein-Gordon theory), and satisfies the canonical commutation relations $(3.1)$.  In coordinate space, the operator
$p_\mu$ in $(3.6)$ is used everywhere in our analysis except where specified, as we shall discuss, in what we shall call the Foldy-Wouthuysen [22] representation, where it takes on the form $-i\hbar {\partial/\partial x^\mu}$.
\par The states $\{\psi_\tau (x)\}$ satisfy the Schr\"odinger-Stueckelberg (see also Schwinger[23] and DeWitt [24]) equation
$$i {\partial \over \partial \tau}\psi_\tau (x)= K \psi_\tau (x), \eqno(3.7)$$
where
$$K = {1 \over 2M}p_\mu g^{\mu \nu}  p_\nu  + V(x), \eqno(3.8)$$
\par The spin of a particle is an essentially quantum mechanical property. In the non-relativistic quantum theory, the lowest non-trivial representation of the rotation group
corresponds to the spin degrees of freedom of the particle. However, for a particle decribed in  the framework of special relativity, the Lorentz group $O(3,1)$ or its covering $SL(2,C)$ acts on the wave function. Wigner [15] showed that representations of $SU(2)$ in the relativistic case, in particular, for spin $1/2$, can be constructed by starting with a particle at rest so that its four-momentum has just one non-zero component, $\pi_0 = m$, where $m$ is the mass of the particle (assumed nonzero; the zero mass case, such as for the photon, must be treated separately). In the four dimensional Minkowsi space this vector lies along the time axis. The elements of the Lorentz group that leave this vector invariant lie in the subgroup $SO(3)$ or its covering $SU(2)$, and therefore provide a representation of spin in that frame. Under a Lorentz boost the vector $(\pi_0,0,0,0)$ may move to a general timelike four vector $\pi_\mu$, but the action of the group remains the same about this new vector, {\it i.e.}, it remains in $SU(2)$. This so-called {\it induced representation } is  then identified by Wigner with the intrinsic spin of the particle. 
\par In the relativistic dynamics of SHP, however, which provides a quantum mechanical Hilbert space for the description of quantum states, this construction is not adequate [12] since the wave functions would transform under a unitary transformation that depends explicitly on the momentum of the state (in momentum representation).  The expectation value of the operator  
$ \xi^\mu= i\hbar {\partial \over \partial \pi_\mu}$ would then not be covariant.
\par This problem was solved in SHP by Arshansky and Horwitz [12] (see also [1]), who constructed an induced representation on a time-like vector $n^\mu$ instead of on the four-momentum. For this vector to transform under the Lorentz group, the generators must have the form 

$$M^{\mu\nu} = \xi^\mu \pi^\nu - \xi^\nu \pi^\mu -i\bigl(n^\mu {\partial \over \partial n_\nu} -n^\nu {\partial \over \partial n_\mu}\bigr)    \eqno(3.9)$$
\par We start, as for the method of Wigner [15], in a frame for which $n^\mu=(1,0,0,0)$. The subgroup of the Lorentz group $O(3,1)$ ($SL(2,C)$) which leaves this $n^\mu$ invariant is $SO(3)$ ($SU(2)$). Under a general Lorentz transformation ${\Lambda^\mu}_\nu$, $n^\mu$ takes on general timelike values in the upper light cone; as shown in [1][12], the wave function then transforms at any point $n$ on the orbit, as
$$ \psi'_{n,\sigma}(\xi) = \psi_{\Lambda^{-1}n,\sigma'}(\Lambda^{-1}\xi) D_{\sigma',\sigma}(\Lambda,n), \eqno(3.10)$$
where $\sigma,\sigma' = \pm 1$ are the spin indices and $D_{\sigma',\sigma}(\Lambda,n)$ is the Wigner $D$-function [15]
$$D((\Lambda,n) = L^{-1}(n)\Lambda L(\Lambda^{-1} n), \eqno(3.11)$$
with $L(n)$ the transformation bringing $(1,0,0,0)$ to $n^\mu$. Since the transformation on the wave function (in $\xi$ or $\pi$ representation) is indepedent of $\pi$, the expectation value of $x^\mu$  is covariant.In discussing the two body case later, we remark that this formulation may be applied to any spin (constructed with Clebsch-Gordan products in the spin space [25]). 
\par We now wish to imbed this structure into the manifold of GR.

\bigskip
\noindent{\it 4. Spin of a Particle in SHPGR}
\bigskip
\par The rather straighforward method we have descrihed above for achieving representations of spin in the framwework of the SHP theory is not adequate for general relativity. Although the orbital part of $M^{\mu\nu}$ can be assumed to transform under local diffeomorphims for $\xi^\mu$ in a small local region, we must explicitly assume that the vector $n^\mu$, whose properties are at our disposal, also has this local covariance property.
\par  There is a theorem
stated in Abraham, Marsden and Ratiu [16] asserting that:
\par Under the $C^r$ map $\varphi$, for $X,Y$ elements of an algebra on an $r$-manifold, $X\rightarrow X'$ and $Y\rightarrow Y'$, $f$ a function on the manifold,
$$ ([X',Y'][f]) \circ \varphi = [X,Y][f\circ \varphi], \eqno(4.1)$$
which establishes an algebraic isomorphism. We give an explicit proof for our construction in the following.

\par In our case 
$$\varphi: \psi_n(\xi) \rightarrow \psi'_{N} (x), \eqno(4.2)$$
where $N$ is the mapping, defined below, of $n$ into the manifold.
We first define the angular momentum in a small neighborhood so that 
the variables $\xi$ can be considered to be very small.
 Under the local diffeomorphism
$$  \varphi: [{M_\xi}^{\mu\nu},{M_\xi}^{\alpha\beta}] \rightarrow [{M_x}^{\mu\nu},{M_x}^{\alpha\beta}].\eqno(4.3)$$  
The Lorentz algebra therefore remains under these local diffeomorphisms, and we can follow the construction of the induced representation for spin, defined in the algebra of $SU(2)$, just as in the flat Minkowski space.
\par In the following, we show explicitly how the theorem of Abraham, Marsden and Ratiu works in our case for the algebra of commutation relations obeyed by
$$M^{\mu\nu} = \xi^\mu \pi^\nu - \xi^\nu \pi^\mu +n^\mu m^\nu - n^\nu m^\mu,\eqno(4.4)$$
where
$$m^\nu =-i{\partial \over \partial n_\nu}. \eqno(4.5)$$
Mapping $(4.4)$ into the manifold by local diffeomorphims, 
$$M^{\mu\nu} = x^\mu p^\nu - x^\nu p^\mu + N^\mu M^\nu -N^\nu M^\mu    \eqno(4.6)$$
remains of the same form, where we assume that the vectors $n^\mu$ and $m^\mu$ transform under local diffeomorphism in the same way (the properties of $n^\mu$ are at our disposal) as the coordinates and momenta, {\it i.e.}, for small $\xi,x$,
$$ d\xi^\mu = {\partial \xi^\mu \over \partial x^\lambda} dx^\lambda \eqno(4.7)$$
and
$$ d\pi_\mu = {\partial x^\lambda \over \partial \xi^\mu} dp_\lambda . \eqno(4.8)$$
We also define
$$ dn^\mu = {\partial \xi^\mu \over \partial x^\lambda} dN^\lambda \eqno(4.9)$$
and
$$ dm_\mu = {\partial x^\lambda \over \partial \xi^\mu} dM_\lambda . \eqno(4.10)$$
To show the consistency of these definitions, we define a Poisson bracket on the variables $\{\xi, n, \pi,m \}$. Generally the motivation for defining the Poisson bracket is in the construction of the $\tau$ derivative of a function $F(\xi,\pi)$ with the use of Hamilton's equations to show that this derivative is given by the Poisson bracket of $F$ with the Hamiltonian $K$. In our case, $K$ generally does depend on $n$, but not, in our discussions so far, on $m$, and therefore  ${\dot n}$ would be zero.  In any case, we may {\it define} a Poisson bracket
$$\eqalign{[A,B]_{PB} &= {\partial A \over \partial\xi^\mu} {\partial B\over \partial \pi_\mu} -
{\partial A \over \partial\pi^\mu} {\partial B\over \partial \xi_\mu}\cr
&+ {\partial A \over \partial n^\mu} {\partial B\over \partial m_\mu} -
{\partial A \over \partial m^\mu} {\partial B\over \partial n_\mu}\cr} \eqno(4.11)$$
\par For brevity, let us define $\{\xi^\mu, n^\mu\} = \zeta^\nu$ and $\{\pi_\mu,m^\mu\} = \eta_\mu$, and the images
$$\zeta^\mu \rightarrow Z^\mu, \ \ \ \eta_\mu \rightarrow E_\mu. \eqno(4.12)$$
With the transformation laws
$${\partial \over \partial \zeta^\mu} =  {\partial x^\lambda \over \partial \xi^\mu} {\partial \over \partial Z^\lambda} \eqno(4.13)$$
and
$${\partial \over \partial \eta_\mu} = {\partial \xi^\mu \over \partial x^\lambda}{\partial \over \partial E_\lambda }, \eqno(4.14)$$
we see that for (summing over both sets of variables)
$$[A,B]_{PB} = {\partial A \over \partial\zeta^\mu} {\partial B\over \partial \eta_\mu} -
{\partial A \over \partial\eta^\mu} {\partial B\over \partial \zeta_\mu} \eqno(4.15)$$
with $(4.13)$
is
$$\eqalign{[A,B]_{PB} &=   {\partial x^\lambda \over \partial \xi^\mu} {\partial A\over \partial Z^\lambda}
{\partial\xi^\mu \over \partial x^\lambda}{\partial B\over \partial E_\lambda }-
 {\partial x^\lambda \over \partial \xi^\mu} {\partial B\over \partial Z^\lambda}
{\partial\xi^\mu \over \partial x^\lambda}{\partial A\over \partial E_\lambda }\cr
& ={\partial A \over \partial Z^\lambda} {\partial B\over \partial E_\lambda} -
{\partial B \over \partial Z^\lambda} {\partial \over \partial E_\lambda}.\cr} \eqno(4.16)$$
\par The Poisson brackets of $x,p$ therefore remain of the form of $\xi,\pi$ and
of $N,M$, the same form as $n,m$. The $N,M$ commutator, as for $n,m$, implies unbounded spectra for $N$ and $M$. We shall be concerned with the constraint, in $N$ representation, that (on the spectrum of $N$) $N^\mu N_\mu = -1$, invariant under the Lorentz algebra generated by $(4.6)$, enabling us to proceed with the program described in $(3.9)-(3.11)$, with $n^\mu$ replaced by $N^\mu$ (also commuting with all dynamical observables).
\par The relations $(4.7)-(4.10)$ are valid for small values of the variables (which can be restricted by the support of the wave function). However, since the algebra is linear it remains valid on the operator level. 
\par We emphasize that the generators we have constructed cannot be simply integrated to form a group on the curved spacetime. However, the spin representations we construct are entirely within the local infinitesimal algebra, sufficient to define spin as a local intrinsic structure of the particle (with generators satisfying the Pauli spin algebra). Constructing higher representations from direct product with Clebsch-Gordan coefficients (coordinate independent) can also be done in the same small neighborhood, as can the composition of spins of different particles [25], as we shall do in our discussion of entanglement.
\par Independently of the coordinate system, $N^\mu$ transforms as a vector under the Lorentz algebra. We may then, as for the flat space, construct a representation of $SU(2)$ as the stability subalgebra of  $N^\mu$ in $SL(2,C)$ . For the definition of the Hilbert space, we remark that there are two
fundamental representations of $SL(2,C)$ which are
inequivalent [26]. Multiplication by, {\it i.e.}, the operator
${\bf \sigma\cdot p}$ of a two dimensional spinor
representing one of these results in an object transforming like the
second representation. Such an operator could be expected to occur in a
dynamical theory, and therefore the state of lowest dimension in
spinor indices of a physical system should contain both
representations (for the rotation subgroup, both of the
fundamental representations yield the same $SU(2)$ matrices up to a
unitary transformation). The defining relation for the fundamental
$SL(2,C)$ matrices is (in the spectral representation of the operator $N_\mu$)
 $$ \Lambda^\dagger \sigma^\mu N_\mu \Lambda = \sigma^\mu
 (\Lambda^{-1} N)_\mu, \eqno(4.17)$$
 where $\sigma^\mu = (\sigma^0, {\bf \sigma})$; $\sigma^0$ is the unit
 $2\times 2$ matrix, and ${\bf \sigma}$  are the Pauli matrices. Since
 the determinant of $\sigma^\mu N_\mu$ is the Lorentz invariant
 ${N^0}^2 - {\bf N}^2$, and the determinant of $\Lambda$ is unity in
 $SL(2,C)$, the transformation represented on the left hand side of
 $(4.17)$ must induce a Lorentz transformation on $N^\mu$  The inequivalent second fundamental
 representation may be constructed by using this defining relation
 with $\sigma^\mu$ replaced by ${\underline\sigma}^\mu\equiv
 (\sigma^0, -{\bf \sigma})$. For every Lorentz transformation $\Lambda$
 acting on $N^\mu$, this defines an $SL(2,C)$ matrix
 $\underline{\Lambda}$ (we use the same symbol for the Lorentz
 transformation on a four-vector as for the corresponding $SL(2,C)$
 matrix acting on the $2$-spinors). 
\par Since, then, both fundamental representations of $SL(2,C)$ should occur
 in the general quantum wave function representing the state of the
 system, the norm in each $N$-sector of the Hilbert space must be defined as
$$ {\cal N}= \int\sqrt{g(x)} d^4x (|{\hat\psi}_N(x)|^2 + |{\hat \phi}_N(x)|^2), \eqno(4.18)$$
where ${\hat \psi}_N$ transforms with the first $SL(2,C)$ and ${\hat
\phi}_N$
 with the second. From the construction of the little group
 $$ D(\Lambda, N) = L^{-1}(N) \Lambda L(\Lambda^{-1} N) \eqno(4.19)$$
 it
follows that $L(n)\psi_N$ transforms with $\Lambda$, and ${\underline
L}(n) \phi_n$ transforms with ${\underline \Lambda}$. Making this
replacement in $(4.18)$, and using the fact, obtained from the
defining relation $(4.17)$, that ${L(N)^\dagger}^{-1}L(N)^{-1}= \mp
\sigma^\mu N_\mu$ and ${{\underline L}(N)^\dagger}^{-1}{\underline
L}(N)^{-1}= \mp{\underline \sigma}^\mu M_\mu$, one finds that
$$ {\cal N}=\mp \int\sqrt{g(x)} d^4x {\bar \psi} _N (x) \gamma\cdot N \psi_N(x),
\eqno(4.20)$$
where $\gamma \cdot N \equiv \gamma^\mu N_\mu$ (for which $(\gamma
\cdot N)^2 = -1)$, and the matrices $\gamma^\mu$ are the Dirac matrices
as defined in the book of Bjorken and Drell [27].  Here, the
four-spinor $\psi_N(x)$  is defined by
$$ \psi_N(x) = { 1 \over \sqrt{ 2}} \left(\matrix{1&1\cr -1 &
1\cr}\right) \left(\matrix{L(N) {\hat \psi}_n(x) \cr {\underline L}(N)
{\hat \phi}_N(x) \cr}\right), \eqno(4.29)$$
and the sign $\mp$   corresponds to $N^\mu$ in the positive or
negative light cone. The wave function then transforms as 
$$ \psi'_N (x) = S(\Lambda) \psi_{\Lambda^{-1}n} (\Lambda^{-1} x)
\eqno(4.30)$$
and $S(\Lambda)$ is a (nonunitary) transformation generated
infinitesimally, as in the standard Dirac theory (see, for example,
Bjorken and Drell [27]), by $\Sigma^{\mu\nu} \equiv { i \over 4}
[\gamma^\mu, \gamma^\nu]$.
\par However, in our formulation, in the dynamics of SHP, we do not obtain the Dirac equation as a factorization of the Klein Gordon equation, but rather a second order equation with Hermition interaction between spin and electromagnetism.
 \par The Dirac operator $\gamma\cdot p$ is not Hermitian ( with $p$ the Hermitian operator defined in $(4.27)$) in the
 (invariant) scalar product associated with the norm $(4.20)$. To construct a Hamiltonian  for the evolution of the wave function consider the Hermitian and anti-Hermitian parts of $\gamma \cdot p$:               
$$\eqalign{ K_L &= { 1\over 2}(\gamma \cdot p + \gamma \cdot N \gamma
\cdot p \gamma \cdot N) = -(p\cdot N) (\gamma \cdot N) \cr 
K_T &= {1\over 2} \gamma^5 (\gamma \cdot p - \gamma\cdot n \gamma\cdot
p \gamma\cdot N) = -2i \gamma^5 (p\cdot K)(\gamma\cdot N),\cr}
\eqno(4.31)$$
where $K^\mu = \Sigma^{\mu\nu}N_\nu $, and we have introduced the
factor $\gamma^5 = i\gamma^0\gamma^1\gamma^2\gamma^3$, which
anticommutes with each $\gamma^\mu$ and has square $-1$ so that $K_T$
is Hermitian and commutes with the Hermitian $K_L$. Since 
$$ K_L^2 = (p\cdot N)^2 \eqno(4.32)$$
and
$$ K_T^2 = p^2 + (p\cdot N)^2, \eqno(4.33)$$
we may consider 
$$ K_T^2 - K_L^2 = p^2 \eqno(4.34)$$
to  pose an eigenvalue problem
analogous to the second order mass eigenvalue condition for the free
Dirac equation (the Klein Gordon condition). For the Stueckelberg
equation of evolution
corresponding to the free particle, we may therefore take
$$ K_0 = {1 \over 2M} (K_T^2 - K_L^2)={1 \over 2M}p^2 . \eqno(4.35)$$
In the presence of electromagnetic interaction, gauge invariance under
a spacetime dependent gauge transformation, the expressions for $K_T$ and $K_L$ given in $(4.31)$,
in gauge covariant form, then imply, in place of $(4.35)$, 
$$K = {1 \over 2M}(p-eA)^2 + {e \over 2M} \Sigma_N^{\mu\nu}
F_{\mu\nu}(x), \eqno(4.36)$$
where 
$$ \Sigma_N^{\mu\nu}= \Sigma^{\mu\nu} + K^\mu N^\nu -K^\nu N^\mu
 \equiv
{i \over 4} [\gamma_N^\mu, \gamma_N^\nu], \eqno(4.37)$$
and the $\gamma_N^\mu$ are defined below in $(4.41)$.
The expression  $(4.36)$ is quite similar to that of the second order Dirac
operator; it is, however, Hermitian and has no direct electric
coupling to the electromagnetic field in the special frame for which
$N^\mu = (1,0,0,0)$ in the minimal coupling model we have given here
(note that in his calculation of the anomalous magnetic moment
Schwinger[23] puts the electric field to zero; a non-zero electric field
would lead to a non-Hermitian term in the standard Dirac propagator,
the inverse of the Klein-Gordon square of the interacting Dirac
equation).  The matrices
$\Sigma_N^{\mu\nu}$ are, in fact, a relativistically covariant form
of the Pauli matrices.
\par To see this, we note that the quantities $K^\mu$ and
$\Sigma_N^{\mu \nu}$ satisfy the commutation relations 
 $$\eqalign{ [K^\mu,K^\nu] &= -i \Sigma_N^{\mu\nu}\cr
[\Sigma_N^{\mu\nu}, K^\lambda] &= -i[(g^{\mu\lambda} + N^\nu N^\lambda)
K^\mu - (g^{\mu\lambda} + N^\mu N^\lambda) K^\nu, \cr
[\Sigma_N^{\mu\nu}, \Sigma_N^{\lambda\sigma}] &= -i[(g^{\nu\lambda} +
N^\nu N^\lambda)\Sigma_n^{\mu\sigma} +(g^{\sigma\mu} + N^\sigma N^\mu)
\Sigma_N^{\lambda\nu} \cr
&-(g^{\mu\lambda} + N^\mu N^\lambda)\Sigma_N^{\nu\sigma} +
(g^{\sigma\nu} + N^\sigma N^\nu) \Sigma_N^{\lambda\nu}].\cr}
\eqno(4.38)$$
Since $K^\mu N_\mu = N_\mu\Sigma_N^{\mu\nu} = 0$, there are only three
independent $K^\mu$ and three $\Sigma_N^{\mu\nu}$. The matrices 
$\Sigma_N^{\mu\nu}$ are a covariant form of the Pauli matrices, and the last of
$(4.38)$ is the Lie algebra of $SU(2)$ in the spacelike surface
orthogonal to $N^\mu$. The three independent $K^\mu$ correspond to
the non-compact part of the algebra which, along with the
$\Sigma_N^{\mu\nu}$ provide a representation of the Lie algebra of the
full Lorentz group.
\par In our construction of the Dirac matrices by studying the spin on the manifold of general relativity in an induced representation, we may see a relation with the work of Fock and Ivanenko [28], discussing the geometrical meaning of the Dirac matrices, and their reference to the ``vierbeins'' of Ricci[29]. 
\par The covariance of this representation follows from
$$ S^{-1} (\Lambda) \Sigma_{\Lambda N}^{\mu\nu}S(\Lambda)
\Lambda_\mu^\lambda \Lambda_\nu^\sigma = \Sigma_N^{\lambda\sigma} . 
\eqno(4.39)$$
\par In the special frame for which  $N^\mu = (1,0,0,0))$,
$\Sigma_N^{i,j}$ become the Pauli matrices ${1\over 2} \sigma^k$
with $(i,j,k)$ cyclic, and $\Sigma_N^{0j} = 0$. In this frame there is
no direct electric interaction with the spin in the minimal coupling
model $(4.37)$.  We remark that there is, however, a natural spin
coupling which becomes pure electric in the special frame, given by
$$ i[K_T,K_L] = -ie \gamma^5 (K^\mu N^\nu - K^\nu N^\mu) F_{\mu
\nu}. \eqno(4.40)$$
It is a simple exercise to show that the value of this commutator
 reduces to $\mp e {\bf \sigma \cdot E}$ in the special frame for
 which $N^\mu (1.0,0,0)$ this operator is Hermitian and would correspond to
 an electric dipole interaction with the spin.
 \par The matrices 
$$ \gamma_N^\mu = \gamma_\lambda \pi^{\lambda \mu}, \eqno(4.41)$$
 where the projection
$$ \pi^{\lambda \mu}= g^{\lambda\mu} + N^\lambda N^\mu ,\eqno(4.42)$$
appearing in $(4.38)$,
 plays an important role in the description of the dynamics in the induced
representation. In $(4.36)$, the existence of projections on each index
in the spin coupling term implies that $F^{\mu\nu}$ can be replaced by
  ${F_N}^{\mu\nu}$ in this term, a tensor projected into the foliation 
subspace.
\bigskip
\noindent{\it 5. Structure of the Wave Function}
\bigskip
 \par Due to the effect of the gravitational field, the structure of the wave function with spin is more complicated than the corresponding wave function with spin in flat Minkowski space. As we have discussed above, the construction of the induced representation implies that the wave function is parametrized by the inducing vector $N^\mu$ in the neighborhood of some point $x$. An infinite number of tangent vectors exist at this point. Any of these can serve as an initial condition in the equation for parallel transport to generate a geodesic curve with the equation for parallel transport, say, for any vector $S^\mu$,
$$ dS_\mu = -{\Gamma^\lambda}_{\mu\nu} dx^\nu S_\lambda, \eqno(5.1)$$
where $dx^\mu$ is along a geodesic curve. Clearly $N^\mu$ as well as, in the associated local frame, the vector $(0,0,0,z)$ must vary along the geodesic curve. Taking $(0,0,0,z)$ as the quantization axis, the spin must then vary along the curve as well. Therefore the wave function contains an {\it ensemble of spins}.
\par Since the geodesic curves cannot cross, the family of geodesics passing through the point $x$, which we shall call $P$, provide a well defined set of spins generated on these geodesic curves. Generally, the set of geodesics  passing through the point $P$ do not densely cover the support of the wave function. These form an equivalence class which we shall label by $N$. In this case, we must choose another point $P'$ and generate another equivalence class, with the constraint that there must be continuity of the spin defined at the common boundaries. If there remain regions not covered by the geodesic curves emanating from $P$ and $P'$, the process must be continued until the support of the wave function is densely filled. The wave function may then be labelled as $ \psi_{N_1,N_2,....} (x)$, according to the equivalence classes, where each of the $N_i$ is associated with an induced representation at the point $P_i$. The norm is defined as in $(4.18)$ or $(4.20)$ with indices summed on the representations at each point. Expectation values of operator functions of $x$ have the same structure. Since the momentum operator moves the wave function along a geodesic, expectation values of operator functions of momentum connect wave functions that are related by parallel transport. The construction of a Fock space would involve tensor products of one particle functions of this type. This topic will be treated elsewhere.
\par It is possible that there are closed geodesic curves in the support of the wave function; these are not included in the general set of open geodesics. Parallel transport on such curves would lead to a multivalued function on the manifold, and we must therefore introduce cuts as in analytic function theory. The structure is somewhat analogous to a type II superconductor or certain configurations in magnetohydrodynamics[30].
\par We work out here explicitly, as an example, the parallel transport of a vector on a closed geodsic curve in the spherically symmetric case (Ludwin and Horwitz studied the bound state solutions for this case [31]) for the Schwarzschild coordinates $t,r,\theta,\varphi$. We assume a geodesic circle at constant $t,\theta$ and $r$ and do the integration over $\varphi$ with measure $d\varphi$. The change in the vector is then
$$ dS_\mu = -{\Gamma^\lambda}_{\mu\varphi} d\varphi S_\lambda. \eqno(5.2)$$
The only nonvanishing components of the connection form that enter are ({\it e.g.}Weinberg[32])
$$\eqalign{{\Gamma^\varphi}_{r\varphi} &= 1/r \cr
  {\Gamma^\varphi}_{\theta\varphi} &= \cot {\theta}\cr
  {\Gamma^\theta}_{\varphi\varphi} &= -\sin{\theta}\cos{\theta} \cr} \eqno(5.3)$$
\par The parallel transport equations can then be written
$$\eqalign{ {dS_r \over d\varphi} &= -{1 \over r} S_\varphi \cr
  {dS_\theta \over d\varphi} &= -\cot{\theta} S_\varphi \cr
  {dS_\varphi \over d\varphi} &= \sin{\theta} \cos{\theta} S_\theta\cr} \eqno(5.4)$$
Due to the non-diagonal structure of the connection form, we see that this sytem is, in fact, second order. At fixed $\theta,r$, differentiating the second and third equations of $(5.4)$,we have
$$\eqalign{{d^2 S_\theta \over d^2\varphi} &= -k^2 S_\theta \cr
  {d^2 S_\varphi \over d^2 \varphi} &= -k^2 S_\varphi, \cr} \eqno(5.5)$$
where $k= |\cos{\theta}|$. The solutions of these oscillator type equations are given by
$$\eqalign{ S_\theta &= A(\theta,r) \cos{k\varphi} + B(\theta,r) \sin{k\varphi}\cr
  S_\varphi &= C(\theta,r) \cos{k\varphi} + D(\theta,r)\sin{k\varphi}.\cr} \eqno(5.6)$$
These solutions determine the equation for $S_r$:
$$ {dS_r \over d\varphi} = - {1 \over r} [C(\theta,r) \cos{k\varphi} + D(\theta,r)\sin{k\varphi}] \eqno(5.7)$$
so that
$$ S_r = - {1 \over kr} [C(\theta,r) \sin{k\varphi} - D(\theta,r)\cos{k\varphi}].\eqno(5.8)$$
\par We must now set initial conditions at $\varphi =0$. 
From Eq. $(5.6)$ we have
$$\eqalign{ A(\theta,r) &= (S_\theta)_0 \cr
C(\theta,r) &= (S_\varphi)_0 \cr} \eqno(5.9)$$
and from $(5.8)$,
$$ D(\theta,r) = r (S_r)_0. \eqno(5.10)$$
\par Since our equations are second order there must be just two independent constants of integration. One can eliminate, say, $B$ and $D$, with initial conditions. 
\par Now, using our solutions $(5.6)$ and $(5.8)$ and the original equations $(5.4)$, we see that
$$\eqalign{ {dS_r \over d\varphi}|_0 &= - {1 \over r} C\cr
             {dS_\theta \over d\varphi}|_0 &= -\cot{\theta} C = kB \cr
             {dS_\varphi \over d\varphi}|_0 &= \sin{\theta}\cos{\theta} A = kD\cr} \eqno(5.11)$$
so that,substituting for $B$ and $D$, we have the solutions
$$\eqalign{S_\theta &= A \cos{k\varphi} - C {\cot{\theta}\over k} \sin{k \varphi}\cr
S_\varphi &= C \cos{k\varphi} + A{\sin{\theta}\cos{\theta}\over k}\sin{k\varphi}\cr
S_r &= - {1 \over kr}(C \sin{k\varphi} - A{\sin{\theta}\cos{\theta}\over k}\cos{k\varphi})\cr} \eqno(5.12)$$
With $A$ and $C$ given by the initial conditions $(5.9)$ this provides the new vector at any $\varphi$; for one traverse, set $\varphi = 2\pi$ at any given $\theta$ for $k =|\cos{\theta}|$. Note that there is no singularity at $\theta =\pi/2$ ($k=0$).
\par Dynamical consequences of this highly complex structure of the wave function in the presence of electromagnetic interaction and in scattering theory[2] will be discussed in a succeeding publication.
 \bigskip
\noindent{\it 6. Entanglement}
\bigskip
\par In this section we discuss the two-body correlations that are associated with an Einstein-Podolsky-Rosen[33] type experiment, in the presence of a gravitational field. 
\par After an initial two-body state is formed, say, by ionization of He, the particles  evolve coherently along geodesic curves,\footnote{*}{Deng {\it et al}[34], for example, have observed quantum correlations over a distance of about $150 \times 10^6$ kilometers.} with motion generated by the free Hamiltonian. We can see this by multiplying the two body wave function by a unitary map which we write formally as
$$ U(p_1,p_2) = \Pi_{\epsilon_1 , \epsilon_2} e^{ i{\epsilon_1}^\mu {p_1}_\mu}e^{ i{\epsilon_2}^\mu {p_2}_\mu} ,\eqno(6.1)$$
for $\epsilon_1,\epsilon_2$ a set of infinitesimal shifts along selected geodesic curves. Then, 
$$ U(p_1,p_2)\psi_{N,\sigma}(x_1,x_2)) = \psi_{N,\sigma}({x_1}' ,{x_2}' ), \eqno(6.2)$$
where ${x_1}' ,{x_2}'$ are points along a classical geodesic curve, ${x_1} ,{x_2}$ translated separately by the sequence of maps in $(6.1)$ corresponding to the free motion of the wave packets in a two body system with given initial conditions.
\par The two body initial state is necessarily formed in the induced representation based on a common $N^\mu$, where we assume the atom is small compared to variations in the gravitational field; otherwise the two $N^\mu$ vectors would be related by parallel transport.
\par The particles then separate along geodesic curves and the $N^\mu$ vector associated with each is parallel transported. The spin correlations then remain as in the initial state. The two body wave function, defined at a given $N^\mu$ therefore maintains the correlation.
\par If we measure the spin of one of the particles in a singlet state in the direction ${\bf n}$, for example (any direction may be chosen), and
find a particle with spin oriented along this direction at some spacetime point A, then at some point B along a geodesic curve connected to A, we are sure to find the spin of the second particle in the $-{\bf n}$ direction of the parallel transported vector, providing us with an EPR [30] situation. This argument is independent of the complexity of the spin content of the wave function, since the spins of the two body state are correlated to the same representation. 
\par In this way, two particles initially in a spin zero state in the spin space, with wave packets moving coherently along geodesic curves, should maintain the EPR correlations in {\it spacetime} (correlations are maintained even for small relative dispacements from the geodesic curve, as discussed in [35]).
 \bigskip
 \noindent{\it 7. Conclusions}
 \bigskip
 \par We have shown that the method of induced representations developed for relativistic quantum mechanics on the Minkowski manifold can be applied as well to the local construction of an induced representation for the spin of a particle on the manifold of general relativity. We have cited the theorem of Abraham, Marsden and Ratiu which assures the existence of an isomorphic mapping, and shown explicitly, using the invariance of Poisson brackets, that the algebraic structure of the induced representation for spin can be mapped into the quantum theory on the manifold of general relativity. As in the relativistic quantum theory on the Minkowski manifold, the wave function may be labelled {\it locally} by a timelike vector, which we have called $N^\mu$,
which is the stability vector for the algebra of the little group that constitutes the spin.
\par The association of the vector $N^\mu$ with the wave function on part of its support can be constructively defined by parallel transport. There are an infinite number of geodesics that pass through a given point $P$, where a timelike vector $N^\mu$ is defined for the local induced representation, but parallel transport along these geodesic curves in general does not reach a dense covering of the support of the wave function unless, of course, the support is very small compared to variations in the gravitational field.
To achieve such a covering, we must introduce additional initial points from which sufficient sets of geodesic curves can be constructed to densely cover the support of the wave function; continuity must then be enforced across boundaries of the geodesically disjoint regions. 
\par Furthermore, in the presence of a gravitational field of non-vanishing curvature, parallel transport of a vector around a closed geodesic curve in spacetime, if such a configuration occurs in the support of the wave function, would bring us to a different vector after the circuit. We must therefore introduce cuts, in analogy to complex function theory, in order for the wave function to be well-defined. The wave function, labelled as $\psi_{N_1,N_2, ,....}(x)$ then constitutes an ensemble of spins.
 \par We have furthermore discussed the two-body spin correlations which can give rise to an entanglement configuration similar to the EPR experiment.
\bigskip
\noindent{\it Acknowledgements}
\bigskip
\par I am grateful to Moshe Chaichian, Asher Yahalom and Yossi Strauss for discussions, and my wife, Ruth, for critically reading the text.
\bigskip

\centerline{\it References}
\frenchspacing
\bigskip
\item{1.}Lawrence Horwitz, {\it Relativistic Quantum  Mechanics},  Fundamental Theories of Physics 180, Springer, Dordrecht (2015a).
\item{2.}L.P. Horwitz, European Physical Journal Plus {\bf 134}, 313 (2019).
\item{3.} L.P. Horwitz, European Physical Journal Plus {\bf 135}, 479 (2020)
\item{4.} E.C.G. Stueckelberg, Helv. Phys. Acta {\bf 14}, 372 (1941).
\item{5.} E.C.G. Stueckelberg, Helv. Phys. Acta {\bf 14},585 (1941).
\item{6.} E.C.G. Stueckelberg, Helv. Phys. Acta {\bf 15}, 23 (1942).
\item{7.} L.P. Horwitz and C. Piron, Helv. Phys. Acta {\bf 66}, 316
(1973).
\item{8.}  Isaac Newton, {\it Philosophia Naturalis
Principia Mathematica},
London 1687.
\item{9.} I.B. Cohen and A. Whitman  {\it The
Principia: Mathematical Principles of Natural Philosophy: A  New
Translation}, University of California Press, Berkeley (1999).
\item{10.}R.I. Arshansky and L.P. Horwitz, Jour. Math. Phys. {\bf 30},
66 (1989a).
\item{11.}R.I. Arshansky and L.P. Horwitz, Jour. Math. Phys. {\bf 30} 380 (1989c).
\item{12.} R.I. Arshansky and L.P.Horwitz, Jour. Math. Phys. {\bf 30}, 213 (1989b).
\item{13.} L.P. Horwitz and R.I. Arshansky, {\it Relativistic Many-Body Theory and Statistical Mechanics}, IOP, Bristol; Morgan \& Claypool, San Rafael (2018).
\item{14.} M.Land, J. Phys. Conf. Ser.  {\bf 845} 012024 (2017).
\item{15.} E.P. Wigner, Annals of Math. {\bf 40}, 149 (1939).
\item{16.}  A. Abraham,J.F. Marsden and T. Ratiu, {\it Manifolds, Tensor Analysis, and Applications}, Springer Verlag, New York (1988).
\item{17.}  P.A.M. Dirac, {\it Quantum Mechanics}, 1st edition ,Oxford University Press, London (1930);3rd edition (1947).
\item{18.} L.P. Horwitz, A. Gershon and M. Schiffer, Found. Phys. {\bf 41},
141 (2010).
\item{19.} A. Gershon and L.P. Horwitz, Jour. Math. Phys. {\bf 50}, 102704 (2009).
\item{20.} Ana Cannas de Silva, {\it Lectures on Symplectic Geometry}, Lecture Notes in Mathematics 1764, Springer (2006).
\item{21.} T.D. Newton and E.Wigner, Rev. Mod. Phys. {\bf 21}, 400 (1949).
\item{22.} L.L. Foldy and S.A. Wouthuysen. Phys. Rev. {\bf 78}, 29 (1950).
\item{23.} J. Schwinger, Phys. Rev. {\bf 82}, 664 (1951).
\item{24.} B.S. DeWitt, Physics Reports {\bf 19}, 295 (1975).
\item{25.} L.P. Horwitz and M. Zellig-Hess, Jour. Math. Phys. {\bf 56}, 092301 (2015b).
\item{26.} H. Boerner, {\it Representations of Groups}, North Holland, Amsterdam (1963).
\item{27.} J.D. Bjorken and S.D. Drell. {\it Relativistic Quantum Mechanics} McGraw Hill, New York (1964).
\item{28.} V. Fock and D.Ivanenko, Zeit. f. Phys. {\bf 54} 798 (1929).
\item{29.} G. Ricci-Curbastro and T. Levi-Civita, {\it M\'ethodes de calcul differential absolu et leurs applications,} Math. Ann. {\bf 54} (1900).
\item{30.} Asher Yahalom, Phys.Lett. A {\bf 377} 1898 (2013).
\item{31.} D.M. Ludwin and L.P. Horwitz, Jour. Math. Phys. {\bf 52} 012303 (2011).
\item{32.} S. Weinberg, {\it Gravitation and Cosmology: Principles and Applications of the General Theory of Relativity}, John Wiley and Sons, New York (1972).

\item{33.} A. Einstein, B. Podolsky, and N. Rosen, Phys. Rev. {\bf 47}, 777 (1935).
\item{34.}Yu-Hao Deng, et al Phys. Rev.Lett. {\bf 123} 080401 (2019).
\item{35.} L.P. Horwitz and R.I. Arshansky, Phys. Lett. A {\bf 382}, 1701 (2018).

\end